\documentclass[12pt]{article}
\usepackage{amsmath}
\usepackage{amssymb}
\usepackage{epsfig}
\usepackage{euscript}
\usepackage{fancybox}
\usepackage{color}

\def\mathswitchr#1{\relax\ifmmode{\mathrm{#1}}\else$\mathrm{#1}$\fi}

\newcommand {\Was}{W\c as}
\newcommand {\KKMC}{\hbox{${\cal KK}$}\ MC}

\newcommand {\pslash}{\hbox{$\not\hbox{\kern-2.3pt $p$}$}}

\newcommand {\dsISR}[2]{{d\sigma}^{\rm ISR (#2)}_{#1}}

\usepackage{cite}

\usepackage{epic}

\def\Was{W\c as}

\def\alf1{ {\alpha\over\pi} }

\begin{document}
\begin{titlepage}
\begin{flushright}
 {\bf BU-HEPP-03/12, UTHEP-03-1201 }\\
{\bf Dec. 2003}\\
\end{flushright}
 
\begin{center}
{\Large Comparisons of Fully Differential Exact Results for ${\cal O}(\alpha)$
Virtual Corrections to Single Hard Bremsstrahlung in $e^+e^-$ Annihilation
at High Energies$^{\dagger}$
}
\end{center}

\vspace{2mm}
\begin{center}
{\bf   C.Glosser$^{a,b,c}$, S. Jadach$^{d,e}$, B.F.L. Ward$^{a}$ and S.A. Yost$^{a}$}\\
\vspace{2mm}
{\em $^a$Department of Physics,\\
  Baylor University, Waco, Texas 76798-7316, USA}\\
{\em $^b$Department of Physics and Astronomy,\\
  The University of Tennessee, Knoxville, Tennessee 37996-1200, USA}\\
{\em $^c$Department of Physics,\\
  Southern Illinois University, Edwardsville, IL 62026-1654, USA}\\
{\em $^d$CERN, Theory Division, CH-1211 Geneva 23, Switzerland,}\\
{\em $^e$Institute of Nuclear Physics,
        ul. Radzikowskiego 152, Krak\'ow, Poland,}
\end{center}


\vspace{5mm}
\begin{center}
{\bf   Abstract}
\end{center}
We present comparisons of the fully differential exact virtual 
correction to the important single hard
bremsstrahlung process in $e^+e^-$ annihilation at high energies
, which is essential for precision studies of the Standard Model
from 1 GeV to 1 TeV, as calculated by two completely independent
methods and groups. We show that the two sets of results are in excellent agreement. Phenomenological implications are discussed.
\vspace{10mm}
\vspace{10mm}
\renewcommand{\baselinestretch}{0.1}
\footnoterule
\noindent
{\footnotesize
\begin{itemize}
\item[${\dagger}$]
Work partly supported 
by the US Department of Energy Contract  DE-FG05-91ER40627 and by 
NATO grant PST.CLG.980342.
\end{itemize}
}

\end{titlepage}

\def\Kmax{K_{\rm max}}\def\ieps{{i\epsilon}}\def\rQCD{{\rm QCD}}
\renewcommand{\theequation}{\arabic{equation}}
\font\fortssbx=cmssbx10 scaled \magstep2
\renewcommand\thepage{}
\parskip.1truein\parindent=20pt\pagenumbering{arabic}\par
Now that the Standard Model electroweak theory has been established
at the one-loop level~\cite{sm,cern1}, the stage is set for
studying its consequences as both signal and background 
to the physics objectives of current and planned high energy
colliding beam devices, from 1 GeV to 1 TeV for the cms
energy in $e^+e^-$ annihilations for example. The attendant per mille
level studies necessitate control of the EW higher order radiative
corrections at least to order ${\cal O}(\alpha^3L^3)$ for the
leading log effects and to the exact ${\cal O}(\alpha^2)$.
One set of the important contributions to the latter exact results
are the virtual corrections to the single hard bremsstrahlung
in $e^+e^-$ annihilations, which therefore have been studied by several
groups~\cite{berends,in:1987,jmwy,hans}. Comparisons of these results
are essential in order to gain some confidence in their correctness
and some facility in their use to confront the SM with precision data.
For example, in the effort to exploit the radiative return from
cms energies in the 1-2 GeV regime to the resonance regime of the
$\pi\pi$ system in the Daphne environment, precision predictions
of the type compared in this paper are essential. Similarly, in order to
use the radiative return from 200 GeV to the Z in the final LEPII data 
analysis for precision EW tests, again precision predictions of 
the type compared in the following for the respective return processes 
are essential.

Indeed, in Refs.~\cite{jmwy}, some of us (S.J., B.F.L.W., S.A.Y.) have
presented comparisons of the results in Refs.~\cite{berends,in:1987,jmwy}
and in general a very good agreement was found. However, if
one looks at the comparisons in Ref.~\cite{jmwy}, one can see that
at the level of the NNLL (next-to-next leading log), there was
a difference in the results that was consistent with the different
levels of ``exactness'' in the calculations. Specifically,
the mass corrections are included in Ref.~\cite{jmwy} in a fully
differential way, whereas in Ref.~\cite{berends} the mass corrections
are included but the photon angular variable is integrated over and
in Ref.~\cite{in:1987} the results are fully differential
but the mass corrections are incomplete. These comparisons therefore
can not really test the NNLL, fully differential results
in Ref.~\cite{jmwy}.

The situation has changed recently with the advent for the results of
Ref.~\cite{hans}. Using a completely independent method and calculation,
the authors in Ref.~\cite{hans} have also achieved a fully differential,
exact ${\cal O}(\alpha^2)$ result for the virtual correction to single hard bremsstrahlung in the initial state radiation in 
high energy $e^+e^-$ annihilation,
with particular emphasis on the 1-2 GeV cms energy regime.
This then affords a detail cross check at the NNLL level of the
corresponding results. These comparisons are the subject of this paper.

We note that, already in Ref.~\cite{hans2}, a preliminary 
indirect comparison of the
results in Refs.~\cite{jmwy,hans} has been reported via comparison of the two
Monte Carlo's PHOKHARA~\cite{hans} and ${\cal KK}$ MC~\cite{kkmc:2001}, 
as these two
Monte Carlo's have the realizations of the 
results in Ref.~\cite{hans} (PHOKHARA)
and Ref.~\cite{jmwy} ( ${\cal KK}$ MC). Agreement at the 
per mille level was found
on selected observables. This is a good basis upon 
which to view the results which follow.

\begin{figure}[ht]
\begin{center}
\epsfig{file=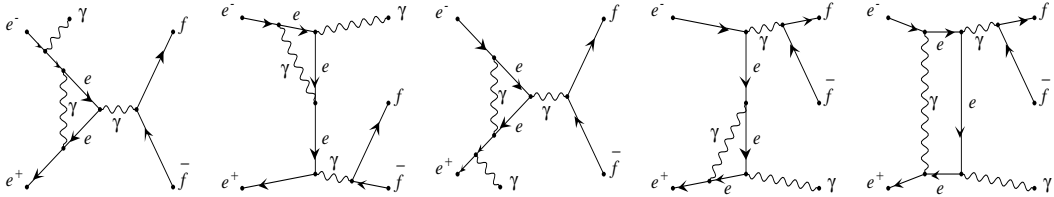,width=140mm}
\end{center}
\caption{\baselineskip=7mm     Feynman graphs for the virtual ${\cal O}(\alpha)$ correction to the 
process $e^+e^-\rightarrow 2f+\gamma$ are illustrated,
where $f\neq e$.}
\label{graph1}
\end{figure}

Specifically, the Feynman graphs under discussion are illustrated
in Fig.~\ref{graph1}.  In Ref.~\cite{jmwy} the respective ISR
matrix element is evaluated using the CALKUL/Xu {\it et al.}/Kleiss-Stirling~\cite{berklei,xuzhangchang,KS} method for the attendant helicity amplitudes using FORM~\cite{form} techniques.
The mass corrections are then added following the methods in
Ref.~\cite{berends1}, after checking that the exact expression
for the mass corrections differs from the result obtained by the
latter methods by terms which vanish as $m_e^2/s\rightarrow 0$
where $m_e$ is the electron mass and s is the squared cms energy. 
In Refs.~\cite{hans}, the same ISR matrix
element's interference with the Born process is calculated
in terms of Lorentz covariants in such a way that the mass effects are treated
exactly. Thus, comparison of the two sets of results gives important
information on the two different methods of calculation and on the
two different treatments of the mass corrections.

For our studies, we will follow the development in Ref.~\cite{jmwy}
and systematically isolate, using the differential cross section 
, first, the the complete cross section, second, the 
size of the LL contributions, third, the size of the NLL contributions
and finally the size of the NNLL contributions. In this way, we
shall see how well the two independent fully differential calculations
with completely different approaches agree. This of course will
reinforce the confidence, whatever it may be, that we 
have in both calculations.

Specifically, for the process in Fig.~\ref{graph1}, where $p_1(p_2)$ is the 
four-momentum of the incoming $e^+(e^-)$, respectively, and $k$ is
that of the emitted real photon, we  focus on the corresponding
virtual correction to the single real bremsstrahlung differential cross section
$\dsISR{1}{1}/dz$ where $\dsISR{1}{1}$ 
is defined in eq.(3.7) of Ref.~\cite{jmwy} and
where $z = s'/s \equiv 1 - v$ when $s$ is the 
squared cms total energy and $s'$ is the squared final state
$f\bar f$ system rest mass. Here, we will average over the initial
fermion spins and sum over the final ones. This cross section has been computed
in both Refs.~\cite{jmwy} and in Refs.~\cite{hans} when we restrict
ourselves, as we do here, to the fully ISR (initial state radiation) 
corrections and processes. Continuing to follow the work in Refs.~\cite{jmwy},
we again note the that IR (infrared) limit of the process is known
and can be determined from the arguments of Yennie, Frautschi and Suura in
Ref.~\cite{yfs}. Thus, with an eye toward the use of the results
in Refs.~\cite{jmwy} in the YFS based CEEX (coherent exclusive exponentiated)
Monte Carlo KK MC~\cite{kkmc:2001}, we work with the cross section
associated with the virtual correction
to the hard photon residual $\bar\beta_1^{(1)}$ as it is defined in
Refs.~\cite{yfs,kkmc:2001} for example. This allows us to focus on the
non-IR singular part of the respective cross sections.

What we find is illustrated in Figs. 2-5, for the
case \hbox{$f\bar f = \mu^- \mu^+$}.
\begin{figure}[ht]
\begin{center}
  		\epsfig{file=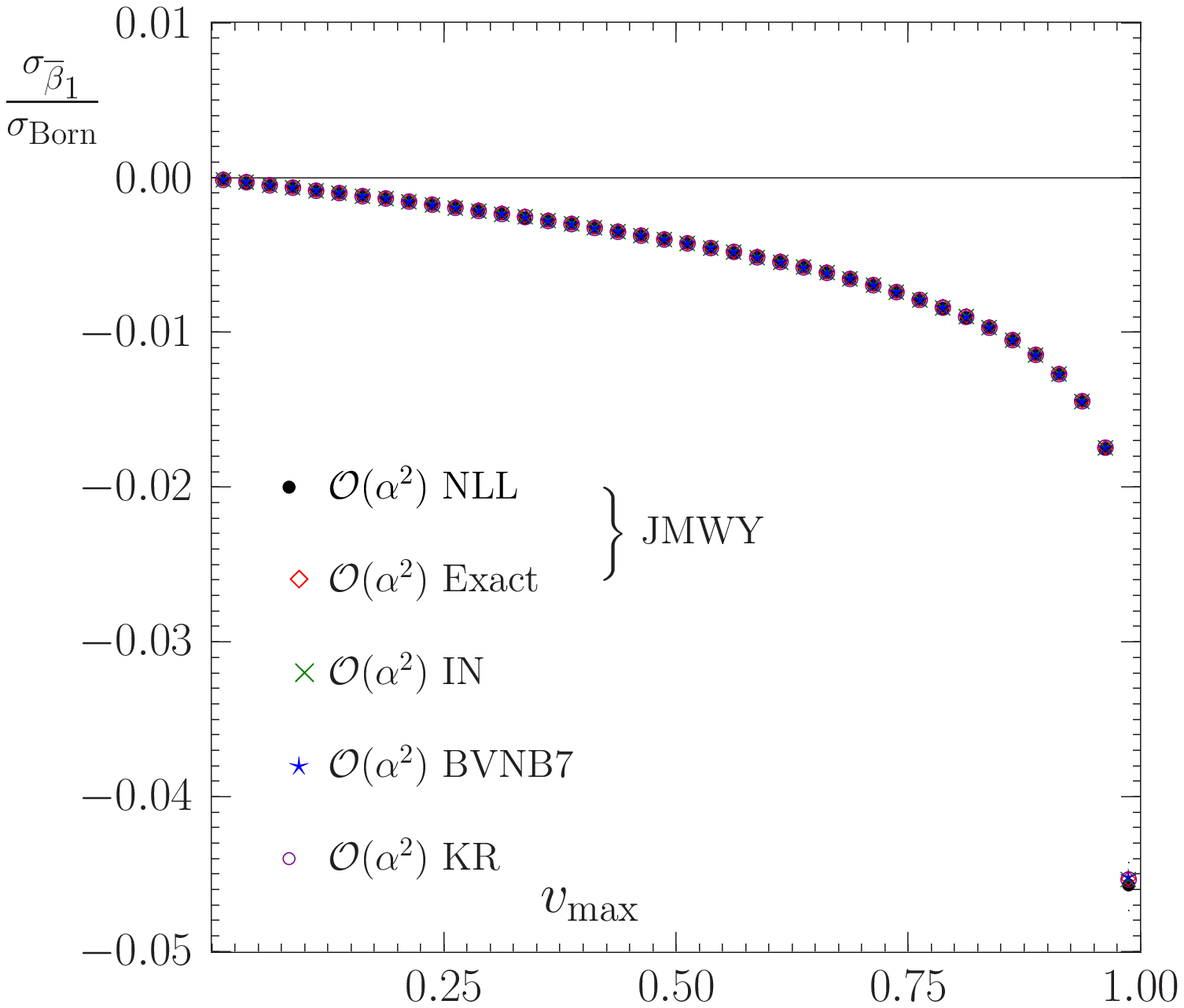,width=16cm}
\end{center}
\vspace{-5mm}
\footnotesize\sf
\caption{
This is the $\bar\beta_1^{(2)}$ distribution for the YFS3ff MC
(YFS3ff is the EEX3 matrix element option of the \KKMC\ in 
Ref.\ \protect\cite{kkmc:2001}), as a function of energy 
cut $v_{\max}$. It is divided by the Born cross-section.    
The exact and NLL results from Ref.\ \protect\cite{jmwy} are compared to
the IN result from Ref.\ \protect\cite{in:1987}, the BVNB result
from Ref.\ \protect\cite{berends}, and the KR result from Ref.~\protect\cite{hans}.
}
\label{fig:Figs-B1}
\end{figure}
In Fig.\ \ref{fig:Figs-B1}, we show the 
complete $\bar\beta_1^{(2)}$ distribution
for our exact result and our NLL approximate results as presented
in Ref.~\cite{jmwy},
the result of Igarashi and Nakazawa {\it et al.}~\cite{in:1987}, 
the result of Berends {\it et al.}~\cite{berends}, and, what is new
here, the exact result of Kuhn and Rodrigo in Ref.~\cite{hans}. 
What we see is that there is a very good general agreement between all of 
these results.  To better assess the difference between them, we plot in 
Fig.\ \ref{fig:Figs-Bdif} the difference between the respective
${\cal O}(\alpha^2)$ and ${\cal O}(\alpha^1)$ results.
Again we see very good agreement between the results.
\begin{figure}[ht]
\begin{center}
  		\epsfig{file=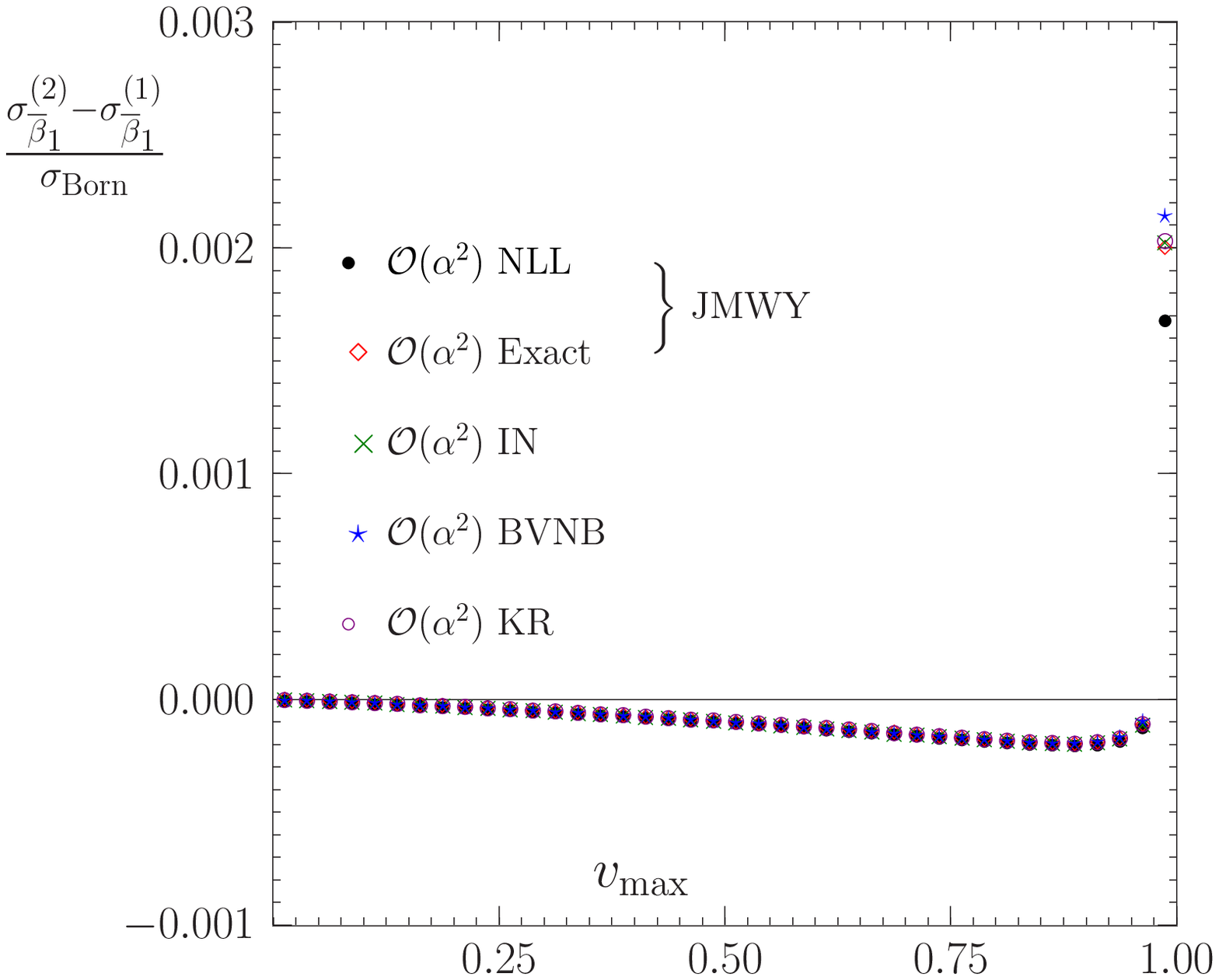,width=16cm}
\end{center}
\vspace{-5mm}
\footnotesize\sf
\caption{
Difference $\bar\beta_1^{(2)} - \bar\beta_1^{(1)}$ for the YFS3ff MC
(the EEX3 option in the \KKMC), as a              
function of the cut $v_{\max}$. It is divided by the Born cross-section.       
The comparisons are the same as in Fig.\ \protect\ref{fig:Figs-B1}.
}
\label{fig:Figs-Bdif}
\end{figure}

\begin{figure}[ht]
\begin{center}
  		\epsfig{file=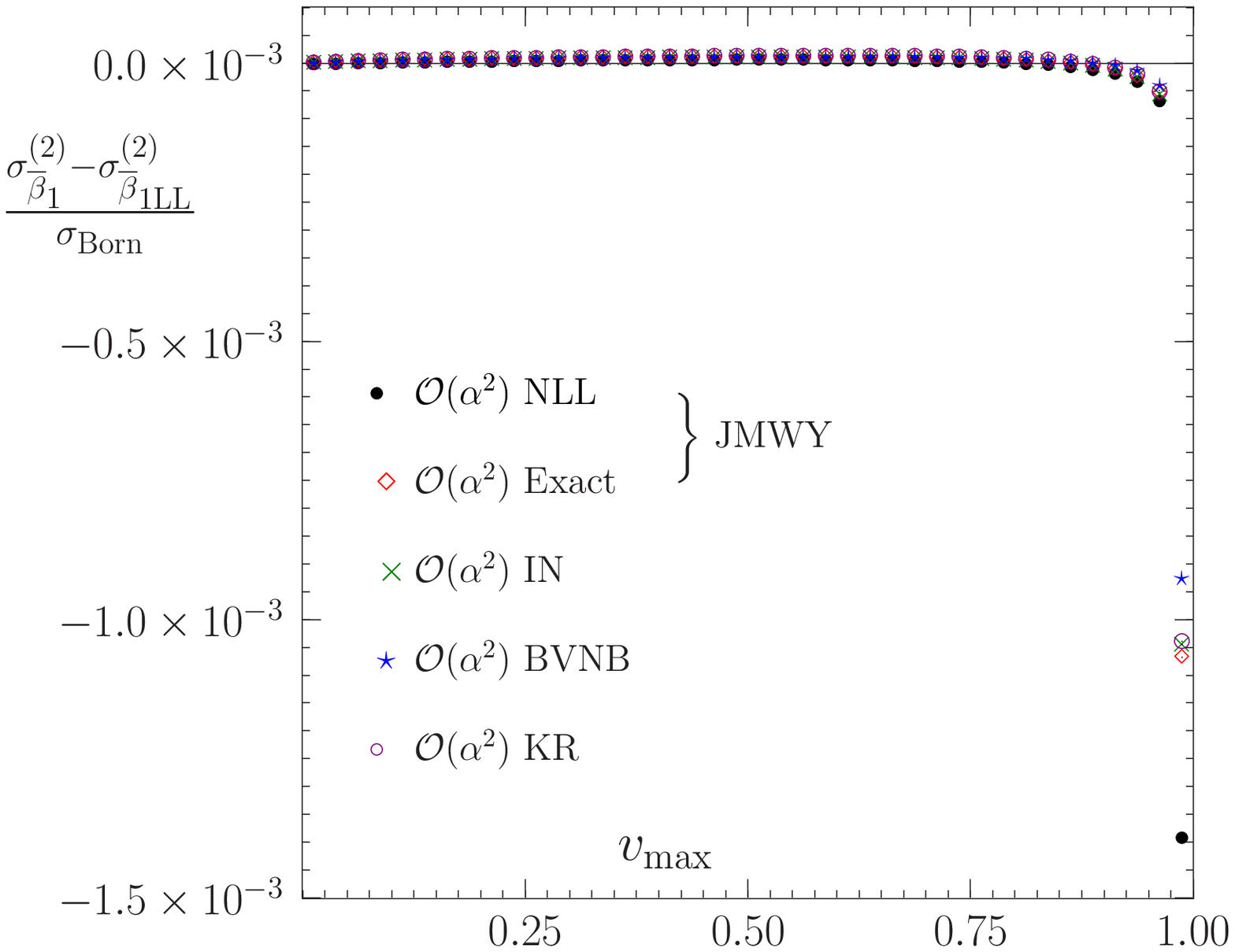,width=16cm}
\end{center}
\vspace{-5mm}
\footnotesize\sf
\caption{
Next-to-leading-log contribution 
$\bar\beta_1^{(2)} -\bar\beta^{(2)}_{1 \rm LL}$ 
for the YFS3ff MC (the EEX3 option of the \KKMC), as a 
function of the cut $v_{\max}$. It is divided by the Born cross-section.       
The comparisons are the same as in Fig.\ \protect\ref{fig:Figs-B1}.
}
\label{fig:Figs-BNLL}
\end{figure}

To isolate the respective predictions for the NLL effect,
we plot in Fig.\ \ref{fig:Figs-BNLL} the respective differences
between our LL ${\cal O}(\alpha^2)$ result from Ref.~\cite{jmwy} and the 
other five results.  The results are still essentially indistinguishable at
this level, except that NLL effects become apparent in the last data point, for
$v = 0.9875$. 

\begin{figure}[ht]
\begin{center}
	\epsfig{file=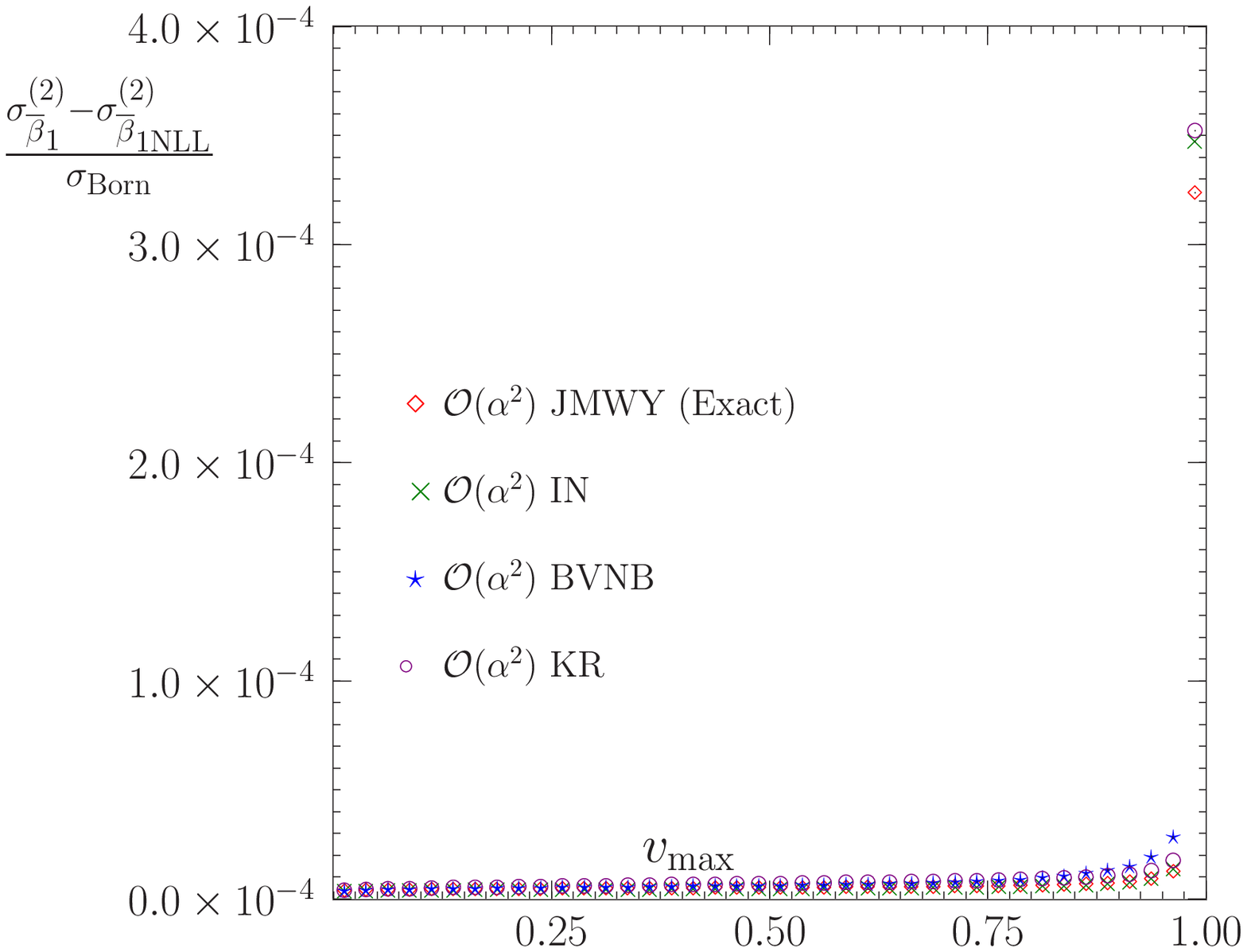,width=16cm} 
\end{center}
\vspace{-5mm}
\footnotesize\sf
\caption{
Sub-NLL contribution $\bar\beta_1^{(2)} - \bar\beta^{(2)}_{1 \rm NLL}$ for the 
YFS3ff MC (the EEX3 option of the \KKMC), as a function of the cut 
$v_{\max}$. It is divided by the Born cross-section.  
The comparisons are the same as in Fig.\ \protect\ref{fig:Figs-B1}.
}
\label{fig:Figs-BNNLL}
\end{figure}

\begin{figure}[ht]
\begin{center}
	\epsfig{file=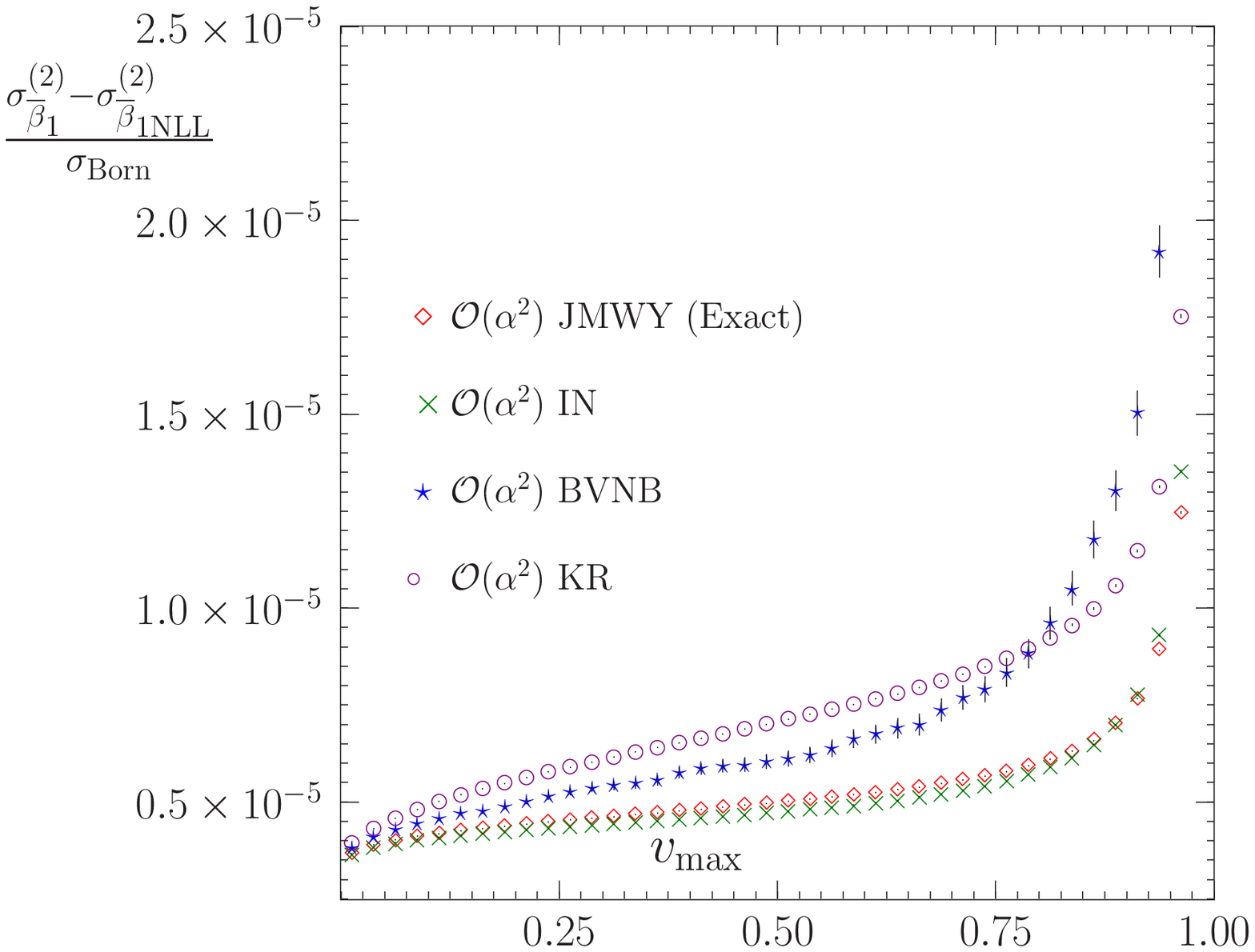,width=16cm} 
\end{center}
\vspace{-5mm}
\footnotesize\sf
\caption{
Sub-NLL contribution $\bar\beta_1^{(2)} - \bar\beta^{(2)}_{1 \rm NLL}$  with
an expanded scale to show clearly the differences in the NNLL results
for cuts up to 0.95.
}
\label{fig:Figs-BNNLL2}
\end{figure}

These comparisons are more evident in the following graphs, 
Fig.\ \ref{fig:Figs-BNNLL} and Fig.\ \ref{fig:Figs-BNNLL2}, where
we isolate the size of the four NNLL results by subtracting
our NLL result from Ref.~\cite{jmwy} from each of the results. 
Fig.\ \ref{fig:Figs-BNNLL2} is identical to Fig.\ \ref{fig:Figs-BNNLL},
except for the scale, which permits a closer comparison of the NLL results
below a cut of 0.95, while omitting some off-scale data points beyond that
cut.  It was 
already established in Ref.~\cite{jmwy} that the expressions of 
Ref.~\cite{in:1987} and Ref.~\cite{berends} agree analytically with our NLL
upon taking a collinear photon limit. This has been checked as well for
the massless limit of the result of Ref.~\cite{hans}. 

In the NNLL comparisons, all of the results agree to within
$0.4\times 10^{-5}$ for cuts below 0.75.  For cuts between 0.75 and .95,
the results agree to within $0.5\times 10^{-5}$, if the result
of Ref.~\cite{berends} is not included, and within $1.1\times 10^{-5}$ if
that result is included. For the last data point, at $v = 0.9875$, the
result of Ref.~\cite{berends} is approximately $1\times 10^{-4}$ greater
than the others, while the remaining results agree to $3\times 10^{-5}$.

These results are consistent with a total precision tag
of $1.5\times 10^{-5}$ for our ${\cal O}(\alpha^2)$
correction $\bar\beta_1^{(2)}$ for an energy cut below $v=0.95$. The NLL effect alone is adequate to within
$1.5\times 10^{-5}$ for cuts below 0.95.
The NLL effect has already been implemented in the
\KKMC\ in Ref.~\cite{kkmc:2001} and the attendant version of KK MC will
be available in the near future~\cite{jad2}.

We have also made the analogous study to Figs. 2-5 for 500 GeV. 
We find very similar results, with the total precision tag of $2\times 10^{-4}$.


In this paper, we have presented new comparisons of exact results for the 
virtual correction to the process $e^+e^-\rightarrow f\bar f +\gamma$
for the ISR. The results from Ref.~\cite{jmwy}
are already in use in 
the \KKMC\ in Ref.~\cite{kkmc:2001} in connection
with the final LEP2 data analysis.

We have compared our results with those in Refs.~\cite{in:1987,berends,hans}
and in general we find very good agreement, both at 200 GeV and at 500 GeV.
The comparisons with the results in Ref.~\cite{hans}, which like our
results in Ref.~\cite{jmwy} are exact but which have the complete mass 
corrections included explicitly whereas our mass corrections are included 
using the approach of Ref.~\cite{berends1}
( which we have shown to differ from the exact mass results by terms
which vanish as $m_e^2/s\rightarrow 0$ ), allow us to lower our
precision tag to $1.5\times 10^{-5}$ for an energy cut below
0.95 compared to what we quoted 
in Ref.~\cite{jmwy}.   
For example, the size of
the NNLL correction is now shown to be at or below the level of $1.5\times 10^{-5}$
for all values of the energy cut parameter up to 0.95. 

We need to stress that a considerable amount of simplification 
of large cancelling terms in the expressions in 
Refs.~\cite{jmwy,hans} was required
to produce the results in this paper. Only after all such cancellations
were found and carried out analytically could the agreement shown
in Figs.~2-6 be realized. The details of this part of the analysis will
appear elsewhere~\cite{jad2}.\par

Our results are
fully differential and are therefore ideally suited for MC 
event generator implementation. This has been done in the \KKMC\ in 
Ref.~\cite{kkmc:2001}. The results from Ref.~\cite{hans}
are also fully differential with mass corrections and have also been fully
implemented into a Monte Carlo event generator~\cite{hans}, albeit
one without YFS exponentiation. It is therefore natural to 
compare the two respective event generators in the context 
of the radiative return 
at $\Phi$ and B-factories. As we noted above, 
a preliminary version of such results 
already appeared in  Ref.~\cite{hans2}, 
where per mille level agreement was found.
The results above indicate that, {\it in principle}, much better agreement 
is possible. Further such comparisons will appear elsewhere~\cite{jad2}.

What one sees from the comparisons above is that 
we now have a firm handle on the precision tag
for an important part of the complete ${\cal O}(\alpha^2)$ corrections
to the $2f$ production process needed for precision
studies of such processes in the final LEP2 data analysis, in the
radiative return at $\Phi$ and B-Factories, and in the
future TESLA/LC physics.

\section*{Acknowledgments}
\label{acknowledgments}

Two of the authors (S.J.\ and B.F.L.W.) would like to thank 
Prof.\ G.\ Altarelli of the CERN TH Div.\ and Prof. D. Schlatter
and the ALEPH, DELPHI, L3 and OPAL Collaborations, respectively, 
for their support and hospitality while this work was completed. 
B.F.L.W.\ would like to thank Prof.\ C.\ Prescott of Group A at SLAC for his 
kind hospitality while this work was in its developmental stages.

\end{document}